\documentclass[letterpaper]{article} 
\usepackage[draft]{aaai25}  
\usepackage{times}  
\usepackage{helvet}  
\usepackage{courier}  
\usepackage[hyphens]{url}  
\usepackage{graphicx} 
\usepackage{natbib}  
\usepackage{caption} 
\usepackage{gensymb}
\usepackage{algpseudocode}
\usepackage{algorithm}       
\usepackage{amsmath}         
\usepackage{subfig}
\usepackage{booktabs}
\usepackage{newfloat}
\usepackage{listings}
\DeclareCaptionStyle{ruled}{labelfont=normalfont,labelsep=colon,strut=off} 
\floatstyle{ruled}
\newfloat{listing}{tb}{lst}{}
\floatname{listing}{Listing}
\title{R2I-rPPG: A Robust Region of Interest Selection for Remote Photoplethysmography to Extract Heart Rate}
\author{
    Sandeep Nagar\textsuperscript{\rm 1}, 
    Mustafa Alam\textsuperscript{\rm 2}, 
    Mark Hasegawa-Johnson\textsuperscript{\rm 3}, 
    David G. Beiser\textsuperscript{\rm 2}, 
    Narendra Ahuja\textsuperscript{\rm 3} 
}
\affiliations{

    \textsuperscript{\rm 1}ML Lab, IIIT Hyderabad, India \\
    \textsuperscript{\rm 2}University of Chicago Medicine \\
    \textsuperscript{\rm 3}University of Illinois Urbana-Champaign 
}

\begin{document}

\maketitle

\begin{abstract}
The COVID-19 pandemic has underscored the need for low-cost, scalable approaches to measuring contactless vital signs, either during initial triage at a healthcare facility or virtual telemedicine visits. Remote photoplethysmography (rPPG) can accurately estimate heart rate (HR) when applied to close-up videos of healthy volunteers in well-lit laboratory settings. However, results from such highly optimized laboratory studies may not be readily translated to healthcare settings. One significant barrier to practical application of rPPG in healthcare is accurate localization of region of interest (ROI). Clinical or telemedicine visits may involve sub-optimal lighting, movement artifacts, variable camera angle, and subject distance. This paper presents an rPPG ROI selection method based on 3D facial landmarks and patient head yaw angle. We then demonstrate robustness of this ROI selection method when coupled to Plane-Orthogonal-to-Skin (POS) rPPG method when applied to videos of patients presenting to an Emergency Department for respiratory complaints. Our primary contributions are twofold: (1) a robust ROI selection framework that adapts to real-world clinical scenarios, and (2)  first unrestricted rPPG dataset collected from emergency ward settings, addressing  critical gaps between controlled laboratory conditions and real-world clinical environments. Our results demonstrate effectiveness of our proposed approach in improving accuracy and robustness of rPPG in a challenging clinical environment.
\end{abstract}

\section{Introduction}
Telemedicine, which delivers medical care through phone and video technology, has been vital for increasing access to healthcare for at-risk populations, especially during the COVID-19 pandemic. It became a key tool for maintaining care while minimizing viral transmission. However, telemedicine has posed challenges for diagnosing and treating patients remotely, particularly in obtaining vital signs like heart rate (HR), which are central to diagnostics. Traditional methods, such as palpation or sensor-based approaches like ECG or pulse oximetry, require patient contact or specialized equipment, which can be costly and inaccessible, especially during a pandemic \cite{beleche2022characteristics}.

\begin{figure}[!ht]
    \centering
    \includegraphics[width=0.99\linewidth]{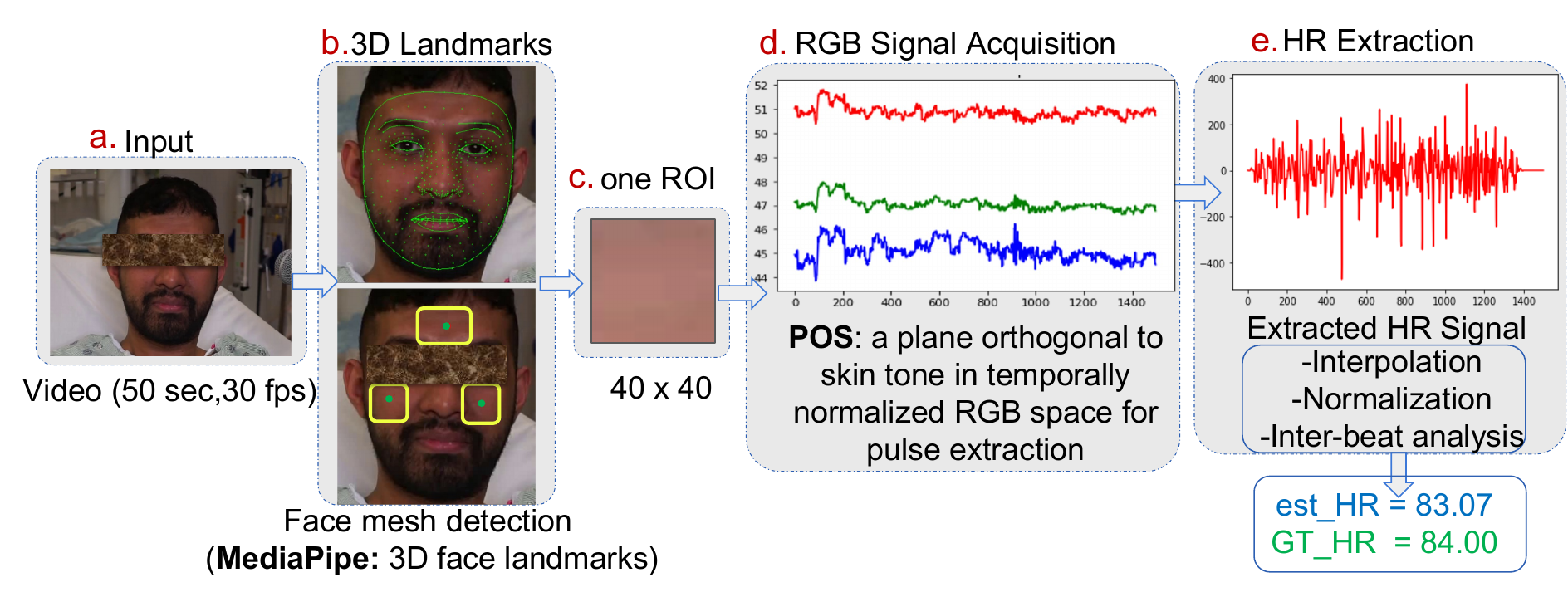}
    \caption{
    Overview of \emph{R2I-rPPG} for real-time heart rate extraction:  (a) input video (b) face detection with 3D landmark localization, (c) ROI definition using landmarks, (d) temporal color averaging over ROIs, (e) POS algorithm application for raw HR signal extraction, and heart rate calculation via interbeat analysis. (fps=frames per second, $ext_{HR}$= Extracted HR, $GT_{HR}$= ground truth HR)}
    \label{fig:flow}
\end{figure}
Non-contact heart rate (HR) extraction using remote photoplethysmography (rPPG) detects periodic micro-color variations from blood flow. The rPPG pipeline involves four stages: 1) extracting a region of interest (ROI) from the video, 2) averaging RGB signals over the ROI to form a 3-D signal, 3) extracting a 1-D PPG signal with minimized noise, and 4) analyzing the PPG signal to estimate HR.

The forehead or cheeks are typically chosen as ROIs due to their high vascularity. For example, method proposed by \citet{verkruysse2008remote} requires user to manually select an ROI through a graphical user interface. Numerous studies have attempted to address artifacts in heart rate detection caused by factors such as head rotation \cite{chen2018video}, facial expressions, illumination variations \cite{lee2022real}, changes in skin tone, motion artifacts, and variable alignment of ROI to face of subject \cite{ zheng2020non}. However, most skin segmentation and tracking algorithms lack standardized methods for selecting and tracking ROIs affected by skin tone variations and head rotation. Most existing automatic ROI detection algorithms are also computationally expensive and vulnerable to facial hair and face mask errors \cite{3Dface}. RGB channels contain the most information about color changes corresponding to blood volume pulse \cite{de2013robust}.  To suppress noise in PPG, \citet{de2013robust} assume a standardized skin color to white-balance in video frames and use chrominance analysis to extract blood volume pulse. Paper by \citet{mcduff2014improvements} project RGB signal to PPG using principal component analysis (PCA), identifying subspace containing most variation due to blood flow. \citet{poh2010advancements} and \citet{mcduff2014improvements} both test independent component analysis (ICA) to compute three maximally independent linear projections of RGB signal. \citet{tsouri2015benefits} demonstrated a generalized blind source separation method, which assumes that signal is a linear mixture of independent color channels and that one is due to heart pulsations. An approach called Plane-Orthogonal-to-Skin (POS), proposed by  \citet{wang2016algorithmic}, exploits property that adding two anti-phase signals with same amplitude cancels out specular distortion. Accordingly, their POS method projects color traces onto plane orthogonal to skin tone in temporally normalized RGB space, where intensity component cancels out. Finally, HR extraction can then be achieved using signal processing methods such as those using autocorrelation \cite{parrish2019autocorrelation}, absolute magnitude difference function \cite{yousefi2012adaptive}, or Fourier power and phase spectra \cite{verkruysse2008remote}. 

In this paper, we propose a novel approach to remote photoplethysmography that focuses on robust detection of informative facial regions using 3D landmarks while accounting for head orientation. Our method uniquely combines adaptive ROI selection with signal enhancement techniques to improve PPG signal quality under realistic conditions. To the best of our knowledge, this is the first work to integrate 3D facial landmarks with dynamic ROI selection and real-time heart rate extraction in clinical settings that encompass head rotation, body motion, and varying illumination conditions.

Our proposed method, R2I-rPPG, presents a comprehensive approach to remote photoplethysmography (rPPG) through a systematic four-stage framework. The framework begins with precise facial feature detection and identification, followed by intelligent ROI selection that adapts to head yaw angle variations using identified facial features. The third stage involves signal extraction from selected ROI utilizing  POS method, while final stage encompasses heart rate extraction through filtered signal analysis. This structured approach enables robust heart rate measurement across varying head positions and lighting conditions, addressing key challenges in remote physiological monitoring.

Our method makes following contributions:
\begin{itemize}
    \item We identify ROI in 3D, using landmarks in a 3D representation of a face.
    \item Our ROI selection algorithm dynamically adapts to head rotation by resorting to a region from one of cheeks if forehead gets occluded.
    \item Our HR extraction algorithm works in real-time on a 2.60GHz CPU with 4GB RAM, and proposed R2I-rPPG HR extraction pipeline is effective under real-world clinical conditions.
    \item We introduce first unrestricted rPPG dataset collected from emergency ward settings, addressing critical gap between controlled laboratory conditions and real-world clinical environments for remote vital sign monitoring.
\end{itemize}

\section{Related Work}\label{sec:related_work}

This section presents an overview of available public datasets, the use of 3D face landmarks for face tracking, ROI selection methods, filtering of rPPG signal obtained from ROI, and HR extraction from rPPG signal.

\subsection{Datasets:}\label{sec:dataset}
Few public datasets are available for rPPG-based HR extraction; many such datasets, like those used by \cite{Kwon2012ValidationOH, poh2010advancements}, are private and include forward-facing close-up views of healthy volunteers in nonclinical environments.

\emph{MAHNOB-HCI:} \cite{soleymani2011multimodal} Although this dataset was created for emotion analysis, it has been adopted for testing rPPG algorithms. In this dataset, $30$ participants ($17$ female and $13$ male, ages between $19$ to $40$ years old) were shown fragments of movies and pictures while monitoring them with six video cameras. Each camera captures a different viewpoint, a head-worn microphone, an eye gaze tracker, and physiological sensors measuring ECG, electroencephalogram, respiration amplitude, and skin temperature.

\emph{VIPL-HR:} This dataset \cite{niu2018vipl} contains $2,378$ visible light videos (VIS) and 752 near-infrared (NIR) videos of $107$ subjects. Moreover, dataset contains head movements, illumination variations, and acquisition device changes, replicating a less constrained scenario for HR extraction. In this dataset, all videos were recorded in a laboratory setting.

\subsection{Tracking ROI:}\label{sec:face}
Several methods exist to detect and track ROI, but a method that can track same ROI in video frame sequence in real-time is necessary for HR extraction.

\emph{The Viola-Jones face detection:} This technique can be employed to automatically detect a subject’s face \cite{viola2001rapid}. The method provides a bounding box coordinate defining subject's face. Implementing face detection at every frame is computationally expensive. Moreover, it causes undesired noise because face's output bounding box is inconsistent between successive frames. 

\emph{Adaptive skin detection:} Skin segmentation is performed using an algorithm proposed by Conaire et al. \cite{4270446}. However, this method is not robust enough to change head position.

\emph{3D facial landmark:} Face localization in a single image is challenging \cite{3D_landmarks} due to ambiguous nature of facial landmarks in a 3D perspective. Li et al. \cite{li2014remote} used 3D facial landmarks to detect faces and then track ROI in video frame sequences.

\emph{MediaPipe Face Mesh:} A robust, real-time 3D-landmark detection method \cite{mediapipe}. It is a lightweight machine-learning-based solution typically used for live augmented reality effects. It employs machine learning to infer 3D facial surfaces.  This method does not track landmarks and detects them independently for each frame, which is more accurate. It is an accurate and robust model that iteratively bootstraps and refines predictions. 

\subsection{ROI selection:}\label{ref:roi}
Selecting a suitable ROI for rPPG-based HR extraction is essential and challenging \cite{fouad2019optimizing}. For PPG, we need skin pixels; to acquire them, we have to track ROI in frame sequence, or we can extract same ROI using 3D landmarks. \citet{lee2022real} uses relative saturation value range to extract skin pixels by converting RGB to HSV color and plotting histogram to get threshold which is not adaptive and may vary subject to subject (e.g., it can be affected by hair, skin color, and head rotation).  In \cite{zheng2020non}, bounding box of left eye is used to find bounding box of forehead, but this method does not work when left eye is not detected. \citet{3Dface} identified left and right cheek ROIs based on face patch visibility. In \cite{3Dface}, 68 3D landmarks require temporal localization and an additional step to make them temporally stable over successive frames. There is still a need for more robust ROI detection, as it is an essential factor in rPPG algorithm's performance over a period of time. Existing methods try to track ROI. To remove limitation of ROI tracking, 3D landmarks can be used to acquire same ROI for each video frame independently.

\emph{Integration of multiple ROI:} HR can be extracted from multiple facial skin regions (or different body parts altogether). In \cite{fouad2019optimizing}, ROI selection using skin segmentation from three different facial regions (forehead, left cheek, and right cheek) is presented. This method used skin segmentation to get ROI, which is not an effective nor efficient method to get skin pixels.  \citet{zheng2020non} observed that having additional regions improves predictions. Due to limited diversity of available datasets, no known physical model relates these three regions (see Figure \ref{fig:selectregions}) per current explorations. To our knowledge, no single skin segmentation method invariant to skin color exists and instead of averaging over all skin regions, we can keep individual skin regions as separate sources.

\subsection{RGB Signal Acquisition:}\label{sec:hr_est}
Several approaches exist to extract HR signal from ROI pixels, including color-based (e.g., rPPG)  \cite{ yang2022assessment} and motion-based (e.g., ballistic motion) \cite{balakrishnan2013detecting} techniques. \citet{balakrishnan2013detecting} presented a motion-based approach, observed tiny head oscillations generated by cardiovascular circulation, and extracted pulse signals from trajectories of numerous recorded features. Due to method's reliance on motion tracking, participants were instructed not to move freely throughout experiment. All current methods for extracting HR from a sequence of frames may be divided into machine-learning techniques and non-machine-learning techniques. Machine learning algorithms are inappropriate for real-time applications because they require extensive training and computation. Due to needed computing power, small devices such as smartphones and edge devices are unsuitable for machine learning approach. In addition, output of machine learning model for real-time applications is biased and erroneous.

\emph{ICA \cite{poh2010advancements}}: This method uses decomposition based on blind source separation to achieve independent components from temporal RGB mixtures. They normalized RGB signals, ignoring that PPG signals induced different known relative amplitudes in individual channels. Therefore, this method is approximation-based and gives approximation of original mixture.

\emph{GREEN \cite{verkruysse2008remote}}: In rPPG, extracting data from green channel is preferred over extracting from red and blue channels, as green channel contains fewer artifacts. This work demonstrated that green signal has highest pulsatility, an intrinsic property of a cardiovascular system, but this requires an additional charge-coupled device (CCD), and this method has two main limitations. First, movement artifacts and, second, reduced signal-to-noise due to CCD-generated noise in recorded pixel values. 

\emph{PCA \cite{lewandowska2011measuring}}: Estimates projected signals using an unsupervised data-driven approach and selects best candidate as output. Essential difference between PCA and ICA is their assumptions concerning relationship, specifically whether two signals are correlated or independent. Therefore, this method does not exploit unique characteristics of skin reflection properties and also loses information.  

\emph{CHROME \cite{de2013robust}}: A chrominance-based method that performs color channel normalization to overcome distortion. This method introduces flexibility when estimating projection direction and reduces sensitivity to prior knowledge used for pulse extraction. CHROME eliminates specular reflection components with a projection. However, it exhibits secular residual in projected signals. 

\emph{Plane-Orthogonal-to-Skin (POS):} A mathematical model incorporating pertinent optical and physiological skin properties to increase our understanding of algorithmic principles behind rPPG. The novelty of this algorithm is in using a plane orthogonal to skin tone in a temporally normalized RGB space. POS requires less accurate knowledge \cite{wang2016algorithmic} of blood volume pulse signature and is more tolerant to distortions. It can be considered a greedy algorithm. This work uses a POS algorithm to extract raw HR signals from three ROIs.

\subsection{HR Extraction:}
The extracted HR signal is subject to noise interference, potentially affecting frequency computation. To address this, filtering techniques are applied to enhance signal and improve signal-to-noise ratio, as detailed in previous studies \cite{benedetto2019remote}. HR, a measure of duration between heartbeats called interbeat interval (IBI), is extracted from filtered signal using IBI analysis, as demonstrated in \cite{aygun2019robust}. Additionally, other studies have utilized fast Fourier transform (FFT) for HR extraction \cite{zhang2018heart}.

\section{R2I-rPPG}\label{sec:method}

The methodology and general structure of our remote HR-measuring techniques are illustrated in Figure \ref{fig:flow}. Our approach utilizes \emph{MediaPipe's} Face Mesh for detecting 3D facial landmarks. We study robust ROI selection method based on yaw angle of head, which is combined with signal filtering methods for HR extraction. 

\begin{figure}[!th]
    \centering
  \includegraphics[width=0.83\linewidth]{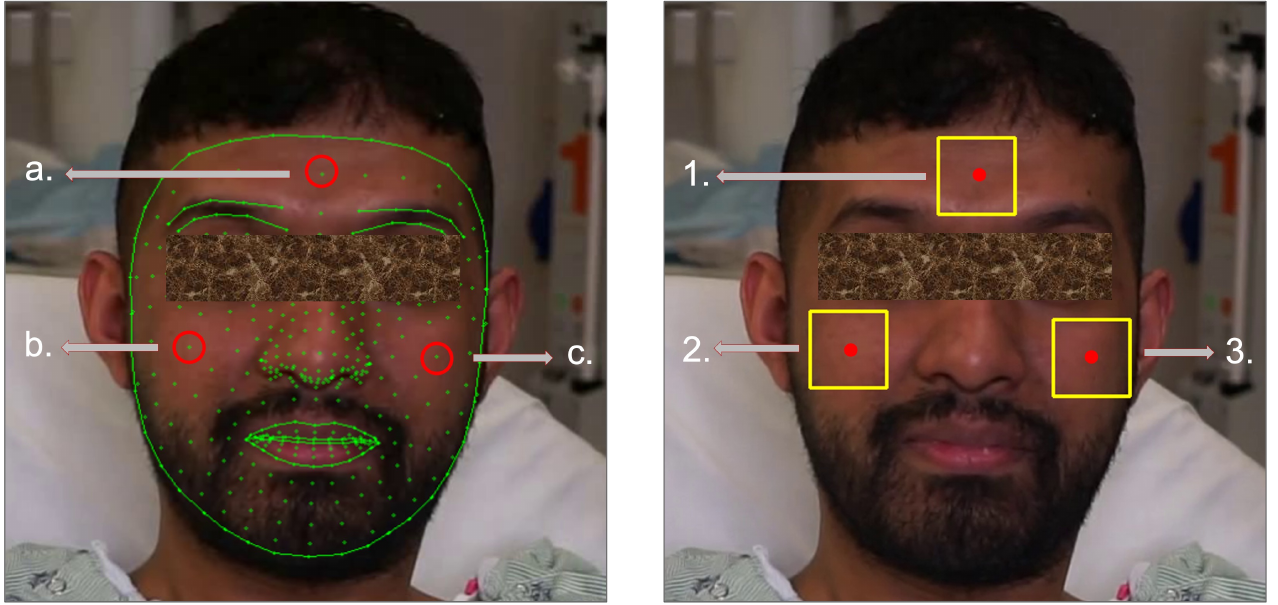}
  \caption{ (a) 3D Face mesh: 468 3D-landmarks (using MediaPipe). (b) Three ROI from 1. forehead center, 2. left cheek, and 3. right cheek (each $40\times40$ centered on respective 3D landmark). ROI's size in pixels, 40x40, is a hyper-parameter and can be set manually based on video's frame size.}
  \label{fig:face_mesh}
\end{figure}

The face detection process begins using \emph{MediaPipe's} Face Mesh, a real-time face detection method. This method provides 468 3-D facial landmarks and is resistant to spatial distortions, appearance distortions, head rotations, and body motion. Figure \ref{fig:face_mesh} illustrates facial landmarks detected by \emph{MediaPipe}. This approach is computationally inexpensive, making it suitable for real-time applications. Once face area is retrieved, ROIs such as cheeks are selected using landmarks and highlighted within face box. Within these ROIs, remote rPPG signal is extracted from pixels. Extracted signal is then subjected to signal extraction techniques, including frequency analysis (Fourier transform) and peak identification  (Inter-beat analysis), to estimate an individual's HR.

\subsection{Face Mesh: 3D-landmarks} Modern face alignment algorithms perform well to automatically detect facial 3D landmarks.  Checking ROI's visibility using head's yaw angle removes requirement for a separate step to determine head's rotation. Our proposed method utilizes facial landmarks to determine ROI, and same face landmarks are recognized independently in each video frame. This removes limitation of ROI tracking. Facial Mesh function of \emph{MediaPipe} extracts 468 3D landmarks from a facial image (see Figure \ref{fig:face_mesh}a). In this work, MediaPipe, a machine learning method, is utilized to infer 3D surface geometry. In our proposed method, we use center of forehead, left cheek, and right cheek in each frame as landmarks for identifying ROI (see Figure \ref{fig:face_mesh}b). \emph{MediaPipe Face Mesh} method returns center of forehead as $151^{st}$ landmark, whereas left cheek is $50^{th}$ and right cheek is $280^{th}$. These landmarks are used as center of ROIs. To overcome restrictions of face tracking or head movement, we use extracted 3D landmarks for each frame independently to locate same ROI across frames.

\subsection{Adaptive ROI selection:}\label{sec:ROI}
Our method leverages multiple facial ROIs (forehead and bilateral cheeks) for robust rPPG signal extraction, selected for their large exposed skin surface area. The ROI selection utilizes 3D facial landmarks to identify fixed $40\times40$ pixel regions at the central forehead and bilateral cheek coordinates. This selection process, formalized in \textbf{Algorithm} \ref{alg:roi}, adapts to varying head poses and visibility conditions.
\begin{algorithm}[!ht]
\caption{\textbf{Adaptive ROI Selection for R2I-rPPG}}\label{alg:roi}
\begin{algorithmic}
\Require Video frame, 3D facial landmarks
\State $ROI\_size \gets 40 \times 40$ pixels
\If{$forehead\_visible$}
    \State $ROI \gets forehead\_region$  \Comment{Primary ROI selection}
\Else
    \If{$yaw\_angle > 15\degree$} \Comment{Head rotation threshold}
        \State $ROI \gets right\_cheek\_region$
    \Else
        \State $ROI \gets left\_cheek\_region$
    \EndIf
\EndIf
\State \Return $ROI$
\end{algorithmic}
\end{algorithm}

\begin{figure}[!ht]
    \centering
    \includegraphics[width=.44\textwidth]{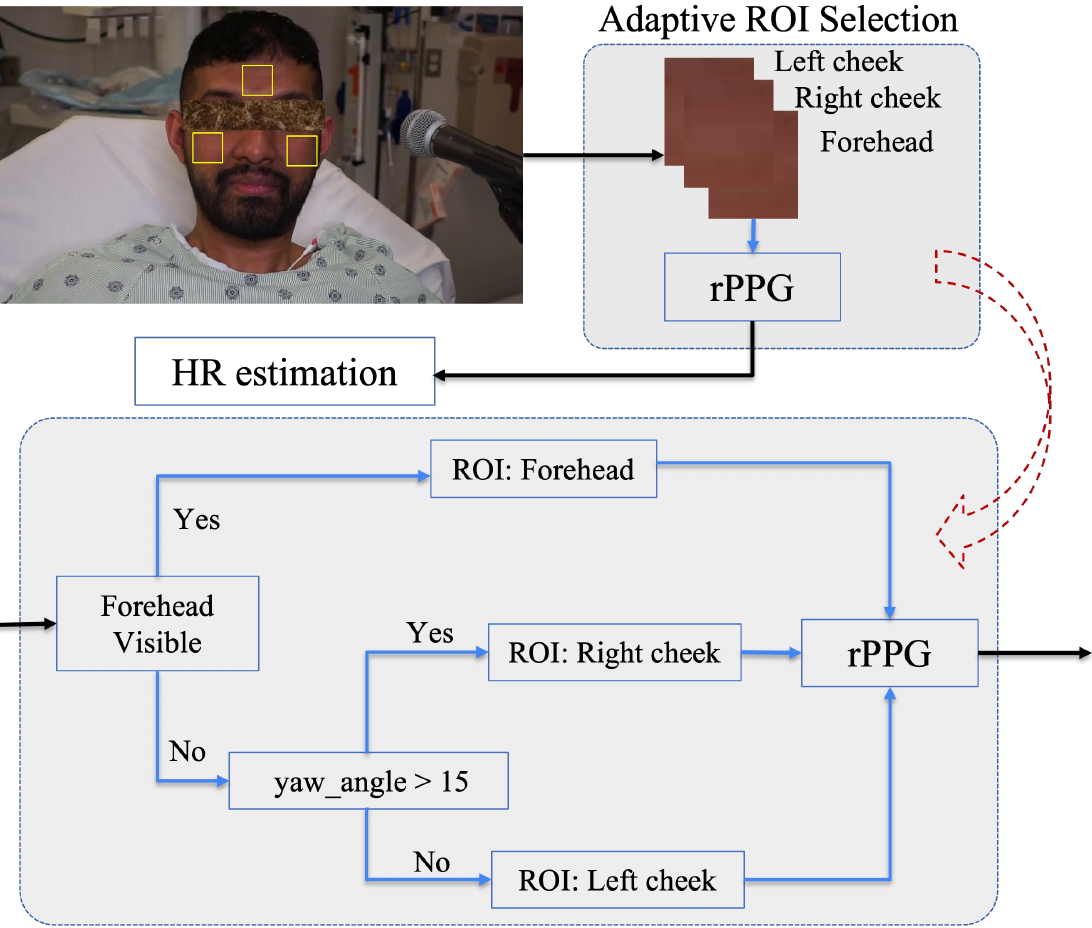}
    \caption{Out of three identifiable ROIs (forehead, right cheek, and left cheek), most appropriate and visible ROI for raw HR signal extraction is selected based on yaw angle.}
    \label{fig:selectregions}
\end{figure}

While the forehead serves as the primary ROI, our system implements an adaptive mechanism for cases of occlusion (by hair, headwear, or accessories). The selection between bilateral cheeks is determined by head yaw angle: the right cheek ROI is utilized when head rotation exceeds 5 degrees leftward, and the left cheek otherwise (Figure \ref{fig:selectregions}). This approach maintains signal quality across different head orientations and occlusion scenarios.

\subsection{HR Extraction:}
\begin{figure}[!ht]
    \centering
    \includegraphics[width=0.48\textwidth]{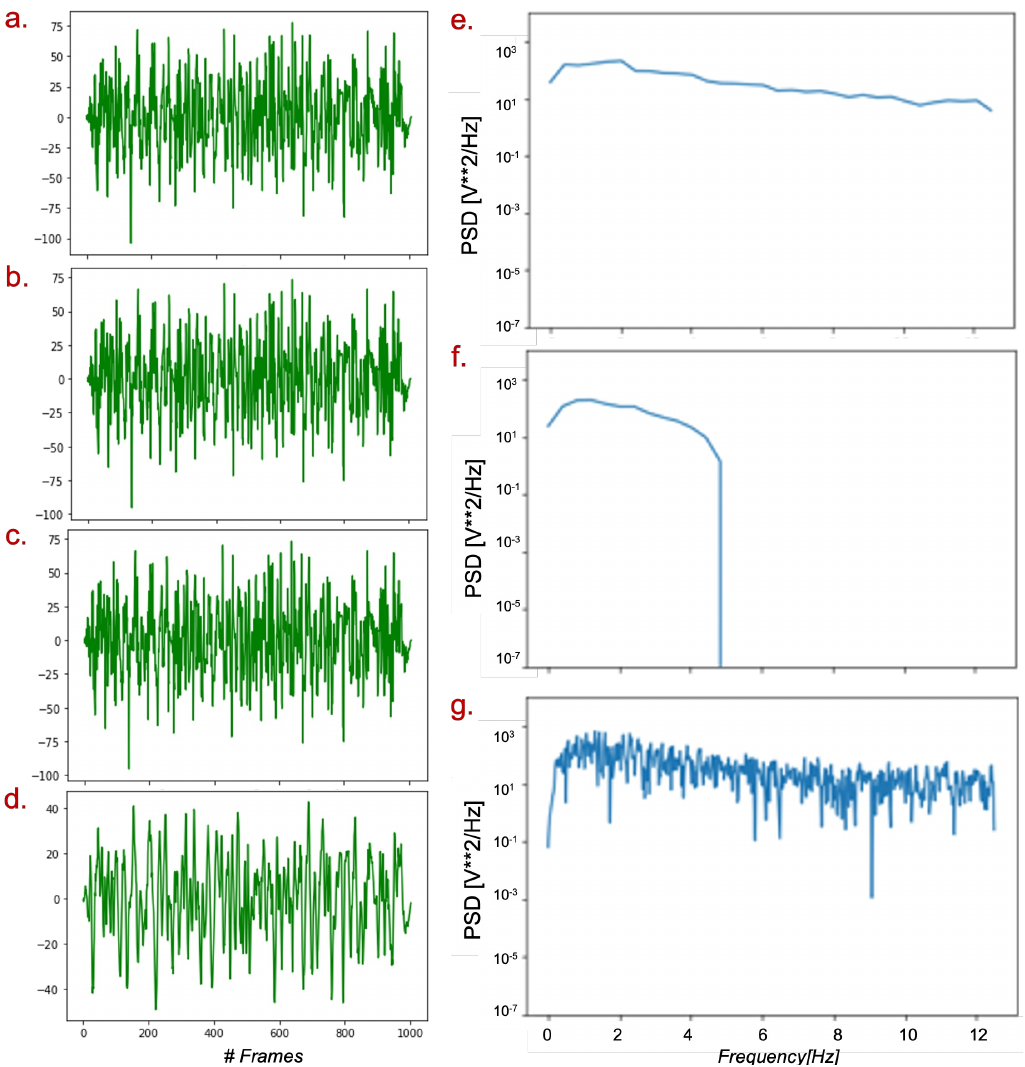}
    \caption{Signal filtering and spectrum analysis. Left: Sequential filtering stages showing (a) raw HR signal, (b) ASF filtered signal, (c) CDF filtered signal, and (d) moving average filtered signal demonstrating noise reduction. Right: Power spectrum estimation using (e) Welch's method, (f) CSD, and (g) interbeat analysis for frequency analysis. PSD: power spectral density.}
    \label{fig:psd-and-filters}
\end{figure}

In this study, ROI is first selected, and then POS algorithm is utilized to extract HR signal from a sequence of frames. As shown in Figure \ref{fig:psd-and-filters}(a), raw extracted HR signal displays fluctuations in intensity within a specific range based on RGB color channels. To effectively extract pulsatile component of HR signal, amplitude selective filtering (ASF) method proposed by  \citet{wang2017amplitude} is applied. ASF is used to select RGB frequency components that fall within assumed pulsatile amplitude range. However, due to differing relative strengths of frequency components caused by cardiac and motion activity in RGB camera outputs, color distortion filtering (CDF) method introduced by \citet{cdf} is employed to improve performance. This method exploits physiological and optical properties of skin reflections. After CDF, a moving average filtering method is applied to extracted signal to remove any remaining random noise while preserving a sharp step response (as shown in Figure \ref{fig:psd-and-filters}d and Equation \ref{eqn:moving_avg}). As observed by comparing Figure \ref{fig:psd-and-filters}c \& d, applying this filter helps increase signal-to-noise ratio by effectively removing noise.

The moving average filter is defined as
\begin{equation}
    y[i] = \frac{1}{M}\sigma_{j=0}^{M-1} x [i + j]
    \label{eqn:moving_avg}
\end{equation}
where $x$ is input signal, $y$ is output signal, and $M$ is number of points in moving average.

To determine HR from a filtered signal, we calculated power spectral density using three methods as shown in Figure \ref{fig:psd-and-filters} (e, f, g). Welch's method resulted in a loss of information for high-frequency values, while Cross Spectral Density (CSD) method resulted in a zero PSD after a certain frequency value. However, Interbeat Interval (IBI) analysis can accurately determine HR within a desired frequency range. Therefore, we used IBI analysis for all videos in this study to calculate HR from a signal. Intervals between consecutive heartbeats are calculated as follows: 

\begin{equation}
    t_{RR,i} = t_n - t_{n-1}
    \label{eqn:interbeat_eqn}
\end{equation}

Where $t_{RR,i}$ is $i^{th}$ cardiac interval in rPPG signal, and $t_{n}$ denotes occurrence of $n^{th}$ peak. Finally, HR is calculated as $HR_w = \frac{1}{ mean IBI_{w}}$, where $mean IBI_{w}$ is mean of IBIs that fall within a time window $w$ and choice of $w$ is a hyper-parameter.

\section{Experiments Settings and Results}\label{sec:experiments}

We implement our proposed method on two datasets: first, our introduced realistic dataset and second, vision for vitals (V4V) \cite{v4v} public baseline dataset. Videos in our new dataset are complex and close to natural conditions. All videos in our dataset were recorded in emergency ward at a hospital. At this time, field of rPPG studies is affected by a lack of datasets, so we believe this new dataset will be a valuable contribution to this field. Below is further elaboration on our dataset.

\subsection{Dataset Collection:}

\begin{figure}[!th]
    \centering
    \includegraphics[width=.46\textwidth]{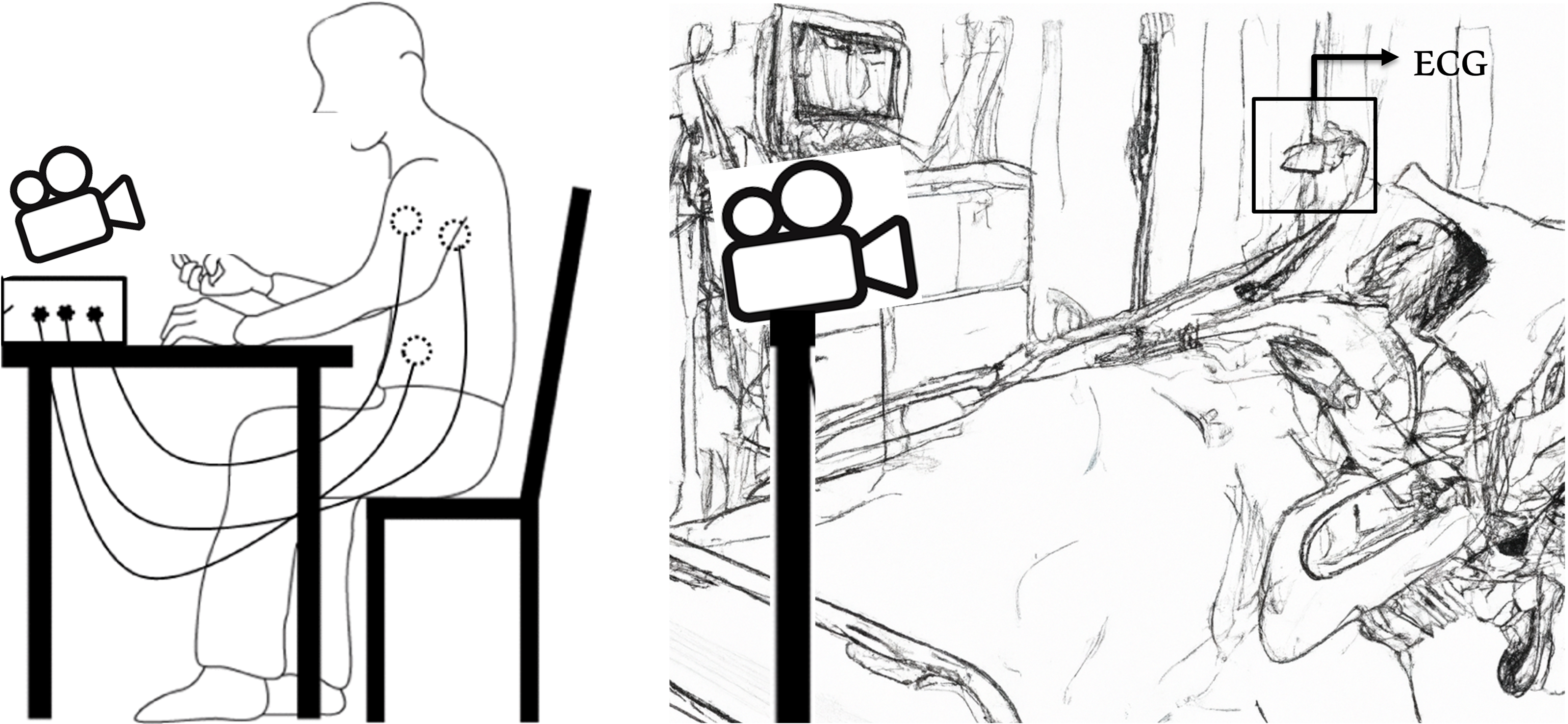}
    \caption{Recording setup comparison: (left) conventional public datasets with controlled settings vs. (right) our unrestricted emergency ward setup allowing natural patient movement and variable camera positions.}
    \label{fig:camera_setting}
\end{figure}

The V4V dataset is more white-skin-dominated. To address lack of diversity in the V4V dataset, a new dataset was created specifically tailored for rPPG, but not limited to it (see video recording setting in Figure \ref{fig:camera_setting}). The purpose of this dataset was to evaluate robustness of proposed HR extraction method and to introduce a new public baseline with diverse face skin colors for rPPG. Videos of subjects lying on a bed were recorded using a fixed camera equipped with ring lights to prevent casting of shadows. videos were recorded at 30 frames per second with a $3840 \times 2160$ resolution using $H.264$ (high profile) coding. subjects were instructed to perform specific movements such as heavy breathing, shifting their body positions, head rotation, and changing facial expressions to introduce variation in dataset.

\subsection{Experimental Settings:} 
The experimental evaluation of our approach was conducted using two distinct datasets. The first dataset, our dataset, comprised videos with durations ranging from 15 to 30 seconds, recorded at 30 frames per second (fps). The second dataset, V4V \cite{v4v}, contained videos spanning 10 to 25 seconds in length, captured at 25 fps. Throughout all experiments, we maintained a consistent ROI with dimensions of $40 \times 40$ pixels. We follow frame-by-frame analysis with continuous forehead visibility verification, enabling dynamic ROI selection based on visibility criteria. For each video in both datasets, a single HR was extracted. To ensure reproducibility and assess computational efficiency, all experiments were performed using a single CPU configuration. This standardized experimental setup facilitated consistent evaluation across both datasets while maintaining computational practicality through focused ROI analysis.

\subsection{Results:}
We evaluate R2I-rPPG on our dataset, down-sampled videos, and the V4V public dataset. To evaluate the accuracy of our method, we employed various statistics commonly used in literature, such as HR error between extracted and ground-truth HR, mean and standard deviation of HR error, root mean squared HR error, and mean of error rate percentage \cite{li2014remote}. 

\begin{figure}[!th]
    \centering
    \includegraphics[width=.41\textwidth]{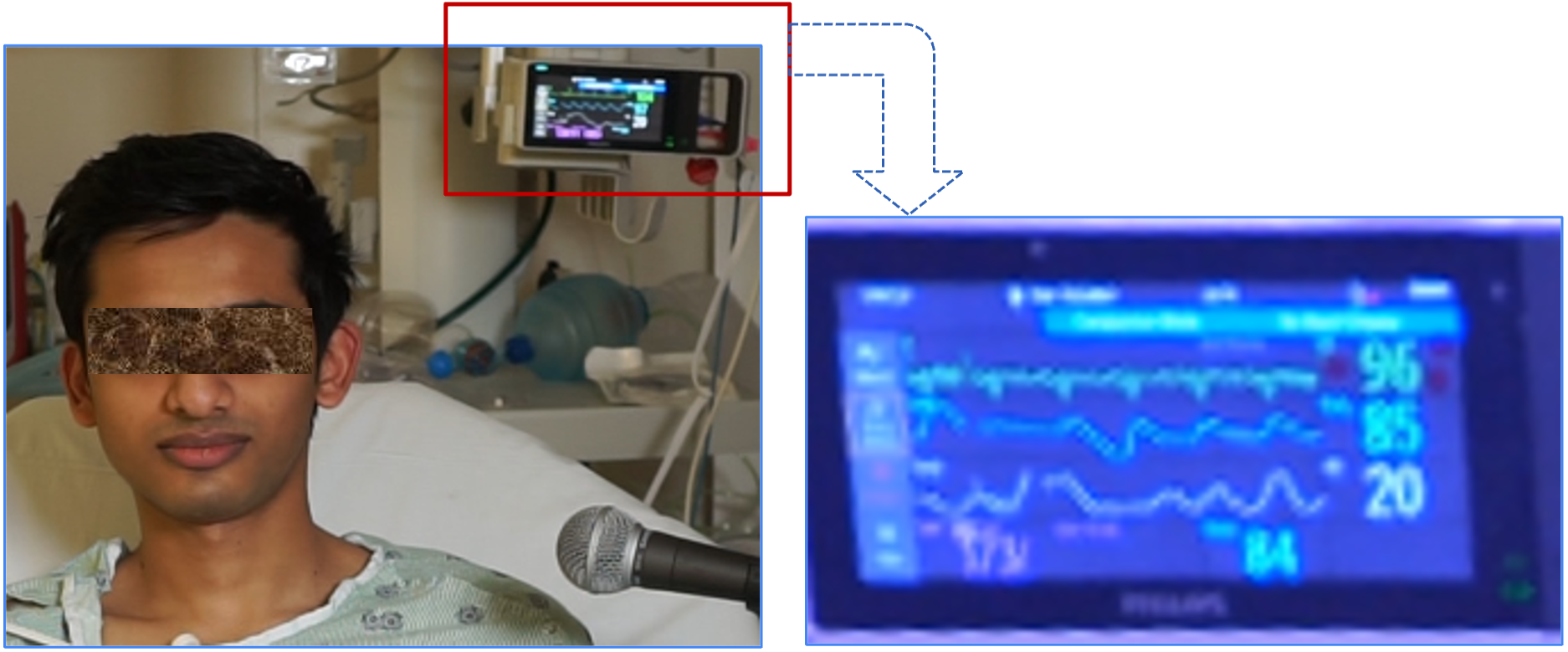}
    \caption{Setup to get ground truth (GT) from video. We use results from ECG readings.}
    \label{fig:ECG_signal}
\end{figure}

\begin{figure*}[!th]
     \centering
     \includegraphics[width=0.71\textwidth]{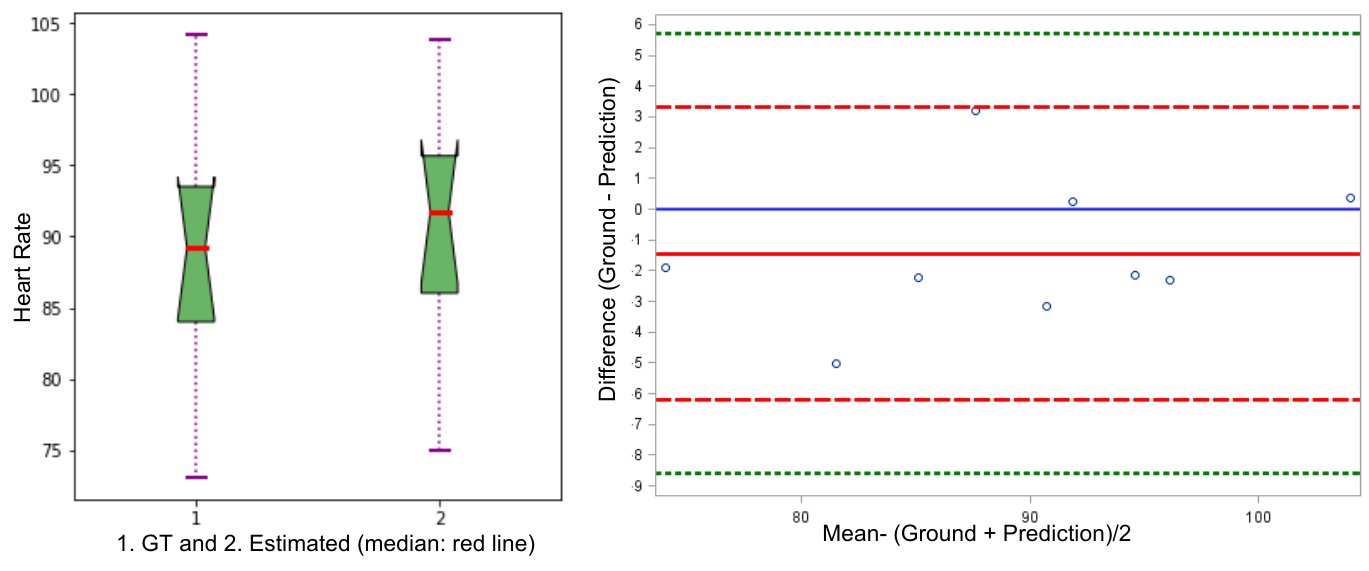}
     \caption{Left: GT and R2I-rPPG extracted HR for nine subjects (Table \ref{tab:table_hr}, \ref{tab:table_hr_2_our_dataset}) with $MAE = 02.28 BPM$. Right: Bland-Altman plot with an average mean difference of $-1.44$.  Solid red line represents mean of difference, dashed red line $\pm 2$ standard deviation (SD) of mean of difference, and dashed green line $\pm 3$ SD of mean of difference.}
     \label{fig:bland_altman}
\end{figure*}

\paragraph{Stationary and Head Motion Condition:}
To validate proposed method, extracted HR values were compared to ground truth (GT) HR  from electrocardiogram (ECG) signal in input video (see Figure \ref{fig:ECG_signal}). For duration of each video sample, we take an average of HR from ECG.  In the scenario where the subject's head remained stationary (Table \ref{tab:table_hr}), measurements from both the forehead and cheek regions showed strong agreement with ground truth values. The forehead measurements demonstrated mean absolute errors ranging from 0.35 to 2.16 BPM from the ground truth. Notably, when selecting the optimal ROI between left and right cheeks (based on yaw angle), the measurements showed comparable or slightly better accuracy, with deviations ranging from 0.84 to 5.37 BPM from ground truth values. In Table \ref{tab:table_hr_2_our_dataset}, we selected five short video clips, ranging in length from 20 to 40 seconds, in which subject's head is not stationary, and extracted HR using our method. When subjects were in motion (Table \ref{tab:table_hr_2_our_dataset}), the measurements showed greater variability, as expected. The forehead measurements deviated from ground truth by 0.18 to 5.03 BPM, while the optimal cheek measurements showed larger variations, ranging from 5.53 to 10.62 BPM.

\begin{table}[!ht]
    \centering
    \begin{tabular}{c|c|c}
     \toprule
       GT & Forehead & Best of L \& R cheek\\
     \bottomrule
      84.0 & 86.21 & 87.01   \\
      92.0 & 91.75 & 92.84   \\
      104.2 & 103.85 & 98.83 \\
      93.5 & 95.66 & 95.12   \\
     \bottomrule
    \end{tabular}
    \caption{Subject's head is not moving: comparison of HR extracted by R2I-rPPG from forehead and best ROI from left or right cheeks, decided by yaw angle.}
    \label{tab:table_hr}
\end{table}

\begin{table}[!ht]
    \centering
    \begin{tabular}{c|c|c}
    \toprule
      GT & Forehead & Best of L \& R cheek\\
    \bottomrule
     89.21 & 86.02 & 94.74  \\
     73.10 & 75.00 & 83.72  \\
     95.00 & 97.30 & 88.06  \\
     89.15 & 92.33 & 81.20  \\
     79.00 & 84.03 & 87.20  \\
    \bottomrule
    \end{tabular}
    \caption{Subject's head is in motion: HR extracted by R2I-rPPG from forehead and optimal ROI between left (L) or right (R) cheek, as determined by yaw angle.}
    \label{tab:table_hr_2_our_dataset}
\end{table}
\paragraph{ROI Analysis for Head Position:}
After down-sampling video quality, we created a few sample videos to validate our proposed method on low-resolution video frames. Using our proposed method, extracted HR for all these video samples. See Table \ref{tab:table_downsampled}; presents extracted HR and GT HR for down-sampled videos. Results obtained in case of down-sampled video reflect that our method works in low-resolution video samples. See Table \ref{tab:table_hr_vital}; we use our proposed method to extract HR for vision for vitals (V4V) \cite{v4v} dataset, which verifies that our method gives significantly better results on public baseline datasets.
\paragraph{Performance on Down-sampled Videos:}
The evaluation of R2I-rPPG on down-sampled videos demonstrated remarkable resilience to reduced video quality. Across eight test subjects (Exp06-01 to Exp17-02), our method maintained high accuracy with minimal deviation from ground truth values: we downsample a set of videos from our dataset and extract HR (see Table \ref{tab:table_downsampled}). This provided a means to evaluate accuracy and reliability of proposed HR extraction method for low-quality videos. To validate proposed method, R2I-rPPG was evaluated using Bland-Altman plots and absolute error (AE) of its estimation: 
$$AE =  |HR_{ext} - HR_{GTs}|$$
where $HR_{ext}$ is extracted HR and $HR_{GT}$ is GT HR form ECG. As in Figure \ref{fig:bland_altman} (left), mean absolute error (MAE) is $02.28 BPM$ for nine subjects (Table-\ref{tab:table_hr} and \ref{tab:table_hr_2_our_dataset}).  Bland-Altman plot obtained from proposed method is depicted see Figure \ref{fig:bland_altman}; this plot compares $HR_{ext}$ and $HR_{GT}$. It can be seen that measurement values all fall inside $98 \%$ bound with $\pm 2 SD$. Our results demonstrate potential of our method in a real-world scenario, where we measure HR from a video of a person talking and rotating their head.

\begin{table}[!ht]
    \centering
    \begin{tabular}{c|c|c|c}
     \toprule
     Subject & GT & R2I-rPPG  & \citeauthor{rouast2017using}    \\
     \bottomrule
     Exp06-01 & 86.70 & 90.45 & 75.01\\
     Exp06-02 & 89.40 & 87.56 & 79.84\\
     Exp06-03 & 90.63 & 87.87 & 78.30\\
     Exp06-04 & 85.30 & 87.37 & 78.80\\
     Exp09-01 & 89.00 & 86.95 & 65.96\\
     Exp10-01 & 91.05 & 89.65 & 74.48\\
     Exp15-02 & 79.01 & 81.48 & 71.34\\
     Exp17-02 & 95.20 & 94.73 & 63.27\\
     \bottomrule
    \end{tabular}
    \caption{HR for down-sampled videos using R2I-rPPG (our) and \citeauthor{rouast2017using}. R2I-rPPG produces comparable results with original high-quality video inputs, even when video quality is reduced.}
    \label{tab:table_downsampled}
\end{table}

\paragraph{Cross-dataset Validation (V4V-Dataset)}
\begin{table}[!ht]
    \centering
    \begin{tabular}{c|c|c|c}
     \toprule
     Subject &  GT & R2I-rPPG & \citeauthor{rouast2017using}\\
     \bottomrule
     F025-T04 & 80.30 & 79.88    &  79.30 \\ 
     F079-T10 & 89.54 & 91.94   &   95.40\\
     F076-T08 & 93.09 & 93.21    &  86.58 \\
     F001-T10 & 110.22 & 108.33 & 104.50 \\
     F017-T05 & 95.04 & 100.09   & 98.57 \\
     F001-T01 & 99.71 & 97.69    & 92.98 \\
     \bottomrule
    \end{tabular}
    \caption{ R2I-rPPG extracted HR for V4V-Dataset, and by \citeauthor{rouast2017using} a real-time method. In all videos, subject's head is always visible. So, R2I-rPPG only considers ROI from forehead. Here, we show that our proposed method works for other baseline datasets.}
    \label{tab:table_hr_vital}
    \vspace{-1em}
\end{table}

These results collectively demonstrate that R2I-rPPG: i) Maintains accuracy across different video quality levels, ii) Shows robust performance across different datasets, iii) Outperforms existing methods, particularly in challenging conditions, iv) Provides reliable measurements regardless of ROI selection, with forehead measurements showing particular stability during motion. The method's consistent performance across these diverse scenarios suggests its potential for real-world applications where video quality and subject movement may vary significantly. 

\section{Conclusion}\label{sec:conclusion}
This work presents R2I-rPPG, a robust remote photoplethysmography framework that leverages both yaw-angle estimation and 3D facial landmark detection for adaptive ROI selection. By introducing 3D landmark-based tracking, our approach overcomes the limitations of traditional face-tracking methods in rPPG systems. The demonstrated real-time performance and low computational overhead make our solution particularly suitable for edge device deployment. Experimental results show strong performance across diverse scenarios, including challenging cases with down-sampled smartphone video inputs. Given the growing importance of remote patient monitoring in healthcare delivery, R2I-rPPG represents a significant step toward accessible, non-invasive vital sign measurement. Future work will focus on expanding the range of extractable physiological parameters and validating the system's efficacy in clinical settings. We believe this research advances the field of camera-based physiological sensing and contributes to the broader goal of democratizing healthcare access through computer vision technologies.

\bibliography{aaai25}
\end{document}